\documentclass[
  journal=largetwo,
  manuscript=article-type,
  year=2020,
  volume=37,
]{cup-journal}

\usepackage{amsmath}
\usepackage{longtable}
\usepackage{amssymb}
\usepackage{hyperref}
\usepackage[nopatch]{microtype}
\usepackage{booktabs}
\usepackage{lscape}
\usepackage{longtable}
\usepackage{subfig}
\usepackage{supertabular,booktabs}
\usepackage{gensymb}
\usepackage{float}
\usepackage{lscape} 
\title{Imaging pulsar census of the Galactic Plane using MWA VCS data}

\author{S.Sett}
\affiliation{International Centre for Radio Astronomy Research, Curtin University, Bentley, WA 6102, Australia}
\alsoaffiliation{CSIRO Astronomy and Space Science, PO Box 76 Epping NSW 1710 Australia}
\email[S. Sett]{284033G@curtin.edu.au}

\author{M.Sokolowski}
\affiliation{International Centre for Radio Astronomy Research, Curtin University, Bentley, WA 6102, Australia}
\author{E.Lenc}
\affiliation{CSIRO Astronomy and Space Science, PO Box 76 Epping NSW 1710 Australia}

\author{N.D.R.Bhat}
\affiliation{International Centre for Radio Astronomy Research, Curtin University, Bentley, WA 6102, Australia}

 %% First letter not capped

\begin{document}

\begin{abstract}

Traditional pulsar surveys have primarily employed time-domain periodicity searches. However, these methods are susceptible to effects like scattering, eclipses and orbital motion. At lower radio frequencies ($\lesssim$ 300\,MHz), factors such as dispersion measure and pulse broadening become more prominent, reducing the detection sensitivity. On the other hand, image domain searches for pulsars are not limited by these effects and can extend the parameter space to regions inaccessible to traditional search techniques. Therefore, we have developed a pipeline to form 1-second full Stokes images from offline correlated high time-resolution data from the Murchison Widefield Array (MWA). This led to the development of image-based methodologies to identify new pulsar candidates. In this paper, we applied these methodologies to perform a low-frequency image-based pulsar census of the Galactic Plane ( 12 MWA observations, covering $\sim$6000 $\rm deg^{\rm 2}$ sky ). This work focuses on the detection of the known pulsar population which were present in the observed region of the sky using both image-based and beamformed methods. This resulted in the detection of 83 known pulsars, with 16 pulsars found only in Stokes I images but not in periodicity searches applied in beamformed data. Notably, for 14 pulsars these are the first reported low-frequency detections. This underscores the importance of image-based searches for pulsars that may be undetectable in time-series data, due to scattering and/or dispersive smearing at low frequencies. This highlights the importance of low-frequency flux density measurements in refining pulsar spectral models and investigating the spectral turnover of pulsars at low frequencies.

\end{abstract}

\section{Introduction}

Even though pulsars were discovered by observing their pulsed emission at a very low radio frequency of 81.5\,MHz \citep{ref:Hewish}, most of the pulsars to date, have been discovered and studied at frequencies $\gtrsim$ 1 GHz. Especially, the large population of Galactic pulsars still remains relatively unexplored at low frequencies ($\lesssim$500\,MHz). The main effects that make pulsar studies at low frequencies more difficult are: (i) scattering \citep[$\propto \nu^{-4}$,][]{ref:scintill}, (ii)  increase in the system temperature due to diffuse Galactic continuum emission \citep[$\propto \nu^{-2.6}$,][]{ref:moz} and (iii) the spectral turnover of pulsars at low frequencies. Despite these effects, multiple Galactic Plane (GP) surveys have been conducted in an effort to discover new pulsars and study the known Galactic pulsar population \citep{ref:pmps,ref:sanidas,ref:gbt,ref:smart1}. With the advancements in instrumentation and computing, studying pulsars at low frequencies is once again coming back to the forefront. Recently upgraded or constructed telescopes such as the Giant Metrewave Radio Telescope \citep[GMRT;][]{ref:swarup,ref:jayanta}, Low-Frequency Array \citep[LOFAR;][]{ref:LOFAR} and the Murchison Widefield Array \citep[MWA;][]{ref:Tingay,ref:Wayth} are contributing to the study of pulsars in the low-frequency regime. 

Studying pulsars at low frequencies will help us better understand the physics of pulsar radio emission and the interstellar medium (ISM). Moreover, given that the vast majority of catalogued pulsars lack reliable flux density measurements below 400 MHz, studying more pulsars at low frequencies is needed to better constraint the models of spectral energy distributions (SEDs) at these frequencies. Studying the radio spectra of pulsars can help in planning surveys of the Galactic pulsar population with the Square Kilometer Array \citep[SKA;][]{ref:keaneska}. 

While the time domain search procedures have been successful in detecting many pulsars in the GP \citep{ref:pmps}, highly scattered or exotic (e.g. sub-millisecond) pulsars are more difficult to detect. Some algorithms that explore a wider parameter space have been developed \citep[for example, acceleration searches][]{ref:Ransom} but they are computationally expensive for all-sky surveys. Continuum images have been considered an effective way to detect known pulsars and pulsar candidates. These image-based efforts have resulted in discoveries of several interesting sources, for example, PSRs J1431-6328 \citep{ref:ASKAP}, J0523-7125 \citep{ref:Wang}. The advantage of searching for pulsars in continuum images is that the detections are unaffected by period, scattering or orbital modulation which is beneficial for GP surveys. Furthermore, similar image-based surveys at low frequencies can be highly effective in overcoming the dispersion and scattering effect that limits the horizon of periodicity searches at low frequencies especially on the GP where the effects of scattering are more prominent. Such image-based GP surveys can be successful in first low-frequency detections of some known pulsars, which may not have been possible to detect at low frequencies by the traditional searches due to scattering becoming more prominent ($\propto \nu^{-4}$). For instance, the first-millisecond pulsar (MSP) was initially detected as an unusually steep and scintillating continuum source and then confirmed by a targeted pulsar search \citep{ref:Backer}. Hence, as this example shows, sensitive image-based searches can lead to a discovery of exotic and/or new classes of pulsars, which were missed by the earlier searches potentially due to being in the parts of parameter space not covered by the traditional searches. The study of pulsars at low frequencies can be used to inform pulsar population studies, which play an important role in estimating the yields of the future surveys \citep{ref:Xue,ref:keaneska}. The MWA is the low-frequency precursor telescope to the SKA and pulsar science is one of the key science goals of the SKA \citep{ref:keaneska}. Therefore, pulsar observations in the same observing environment and at the similar frequencies are necessary to prepare for pulsar science with the SKA-Low. 

In this paper, we present the image-based GP pulsar census using MWA Phase II archival observations. Section 2 describes the main science goal and motivation for the census. Section 3 describes the the observations and the data processing details of this work. Section 4 summarises the results and discusses the implications of this work on the future. Section 5 focuses on summarising the work done as part of this paper and discusses the importance of image-based pulsar candidate search techniques for future surveys and instruments.

\section{Main science goals and motivation}

Very few surveys of the Southern sky such as the GLEAM survey covering the entire Southern radio sky at frequencies between 72 and 231 MHz have been done \citep{ref:gl}. The drift scan observations of the GLEAM survey were able to reach a sensitivity of 10 mJy/beam, producing a catalogue of radio sources that can be used for further discoveries \citep{ref:nhw}. The next generation of the GLEAM survey, GLEAM-X (Ross et al, 2023 in prep) is currently ongoing and is expected to reach down to a sensitivity of 1-2 mJy/beam. 

These surveys have also been utilised to study the pulsar population. Studies of the low-frequency spectral energy distribution of pulsars using the continuum images from the GLEAM survey were realised for 60 radio pulsars \citep{ref:Murphy}. Their analysis provided reliable flux density measurements and helped in improving the spectral modelling of pulsars. Similar studies were performed to explore the variability of pulsars by \citet{ref:Bell} and circular polarisation of pulsars by \citet{ref:Lenc2017}. The method of detecting pulsars using the ISM and variance imaging was also investigated by \citet{ref:Dai}. 33 pulsars were also detected in MWA images as linear polarised sources as part of the POlarised GLEAM Survey (POGS), 11 of which POGS was the first image-plane detection \citep{ref:pogs1,ref:pogs2}. An initial census of Southern pulsars with the MWA Voltage Capture System (MWA VCS) data using incoherent beamforming and searching was performed by \citet{ref:Xue}. Their work also resulted in the first low-frequency detections of 10 pulsars and forecasted that a survey with SKA-Low could potentially detect around 9400 pulsars. The currently ongoing large pulsar survey with the MWA, the Southern-sky MWA Rapid Two-metre (SMART) pulsar survey \citep{ref:smart1,ref:smart2} has discovered 4 new pulsars and is expected to detect many more ($\sim$300 new pulsars after full processing). The SMART survey would also provide a complete census of the known Southern sky population of pulsars (in the time domain) and will be beneficial in informing future pulsar surveys with the SKA-Low.

With the SKA under construction, there is an increased need to study and understand the pulsar population. This work is the first attempt to perform an image-based pulsar census of the dense region of the GP using MWA VCS Phase II data. The expected detectable pulsar population is guided by the current knowledge of the known pulsar population. Therefore it is important to explore all new avenues and refine our knowledge of the known pulsar population. The presented work contributes to this knowledge by detecting some of the pulsars for the first time at frequencies lower than 300 MHz. More detection of pulsars, especially at low frequencies will be useful to address some of the broader questions surrounding the neutron-star population. This work also explores the new parameter space available to image-based pulsar search strategies and provides insights into the efficacy of such methodologies. 

An underlying goal of this work is to provide better and more reliable flux density measurements leading to improved spectral modelling for pulsars that do not have any low-frequency flux density measurements. Finally, conducting a full Galactic census of pulsars is a high-priority science objective for the SKA \citep{ref:keaneska} and this work will also serve as a reference survey for future deeper imaging surveys at low frequencies such as those planned with the SKA-Low. The initial success of this work is the first step towards the greater goal of detecting new pulsars using low-frequency image-based pulsar searches. As the MWA is the official low-frequency precursor for the SKA-Low, the lessons learned from this work will be useful in informing future SKA-Low image-based pulsar surveys.

\section{Observations and data processing}

The MWA is a low-frequency precursor telescope to the SKA, located at the Murchison Radio-astronomy Observatory (MRO) in Western Australia. It operates in the frequency range of 70-300 MHz and is able to access the entire Southern sky. The low-frequency range and the wide field-of-view (FoV), makes it complementary both to similar low-frequency telescopes in the Northern Hemisphere, such as the Low Frequency Array \citep[LOFAR;][]{ref:LOFAR}, as well as high-frequency telescopes in the Southern Hemisphere, such as the Parkes Radio Telescope (Murriyang). The Phase I of the MWA consisted of 128 tiles with a maximum baseline of $\sim$ 3km \citep{ref:Tingay}. In 2018, the MWA was upgraded to phase II with a maximum baseline of 5.3 km and 256 tiles (small 4$\times$4 aperture arrays with dual-polarisation dipoles). The Phase II upgrade increased the angular resolution by a factor of $\sim$ 2 and the sensitivity by a factor of $\sim$ 4 as a result of reduction in the classical and sidelobe confusion \citep{ref:Wayth}. It was initially designed as an imaging telescope, requiring only time-averaged tile cross-correlation products ("visibilities"). However, it was eventually upgraded to enable the capture of raw complex voltages from each tile with the development of the VCS \citep{ref:Tremblay}. The VCS captures high-time and frequency resolution voltage data (100 $\mu s$ / 10 kHz), which provided flexibility and the opportunity to process data in multiple ways. Since pulsar flux densities at low-frequencies can vary significantly from day to day, being able to find pulsar candidates in images formed from the MWA VCS data, and investigate them by beamforming the very same data is a very powerful technique unique to the MWA. The pipeline developed for this kind of processing was described in \citet{ref:Sett}. 

\begin{figure*}[hbt!]
    \centering
    \includegraphics[width=\textwidth]{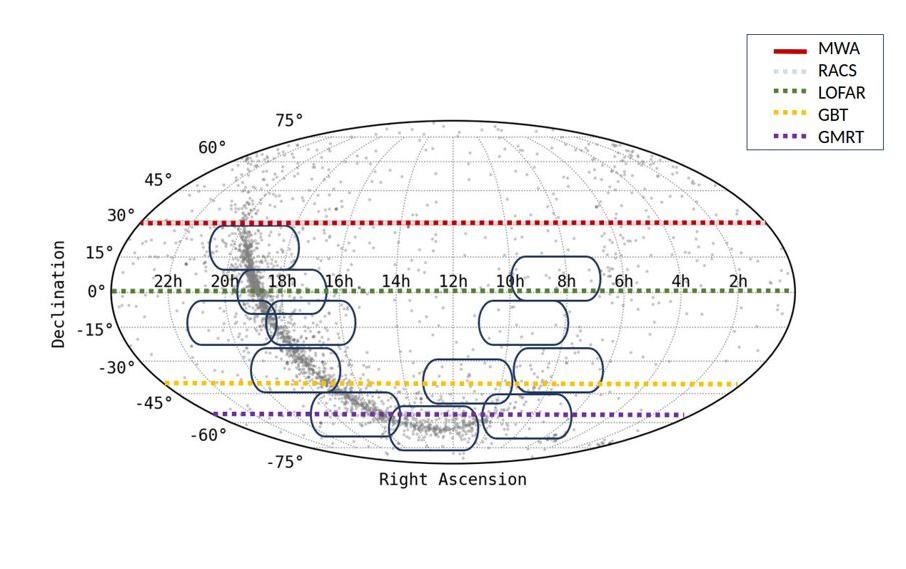}
    \caption[MWA VCS observations processed as part of this work]{The MWA VCS observations processed as part of this work are shown in blue and the details of the observations can be found in Table \ref{table:obs}. The grey points indicate the known pulsars recorded in the ATNF catalogue \citep{ref:ATNF}. The red bold line is the part of the sky available to the MWA and the dotted green, yellow and purple line shows the sky that is visible to LOFAR, GBT and GMRT respectively.}
    \label{fig:obs}
\end{figure*}

The current work uses 12 MWA VCS Phase II observations as shown in Figure \ref{fig:obs}. Their duration ranges between 30 minutes to 90 minutes with a central frequency of 184 MHz. The observations are labelled from A to L and their details are given in Table \ref{table:obs}. The total amount of data processed as part of this work is $\sim$ 450 TB and corresponds to 6000 $\rm deg^{2}$ of the sky. These observations are used to form full Stokes images using the procedure described in Section \ref{subsec:imaging}, which leads to the detection of known pulsars and identification of new pulsar candidates. The mean RMS reached for the Stokes I images of the 12 observations ranges between $\sim$ 5 mJy/beam and 8 mJy/beam. 

\begin{table}[h]
\begin{center}
\resizebox{\columnwidth}{!}{%
\begin{tabular}{ccccccc}
\begin{tabular}[c]{@{}l@{}}\textbf{Obs}\\ \textbf{Name}\end{tabular} & \textbf{OBSID}      & \begin{tabular}[c]{@{}l@{}}\textbf{Duration}\\ \textbf{(s)}\end{tabular} & \begin{tabular}[c]{@{}l@{}}\textbf{Mean RMS}\\ \textbf{(mJy/beam)}\end{tabular} & \boldmath{$\rm N_{\rm src}$} & \boldmath{$\rm N_{\rm psr}$} & \boldmath{$\rm N_{\rm 3db}$}\\
\hline \hline
A                                                  & 1286617815 & 1800                                                   & 8           & 1094                                                  & 4   & 23                                                                   \\
B                                                  & 1284634952 & 5400                                                   & 5           & 9490                                                 & 7     & 78                                                                \\
C                                                  & 1285498952 & 5400                                                   & 5            & 9459                                                 & 7     & 86                                                                 \\
D                                                  & 1275758864 & 5400                                                   & 6            &4296                                                 & 6       & 102                                                               \\
E                                                  & 1206977296 & 3600                                                   & 6             &4126                                                & 6        & 110                                                              \\
F                                                  & 1200918320 & 3600                                                   & 5             &5388                                                & 8         & 123                                                             \\
G                                                  & 1244457047 & 5400                                                   & 5              & 9453                                               & 12       & 150                                                              \\
H                                                  & 1195038136 & 3600                                                   & 6            & 1002                                                 & 7         & 130                                                             \\
I                                                  & 1206201016 & 3600                                                  & 5              & 1035                                               & 8          & 92                                                            \\
J                                                  & 1287796112 & 1800                                                   & 5             & 1028                                                & 6          & 54                                                            \\
K                                                  & 1240826896 & 3600                                                   & 5             & 3626                                                & 9          & 41                                                            \\
L                                                  & 1239460849 & 1800                                                   & 8              & 968                                               & 3      & 11 \\   \hline
\end{tabular}% 

}
\caption[Details of 12 observations]{The details of the 12 observations processed as part of this work, duration, mean RMS achieved for the Stokes I image. The survey covered the dense region of the Galactic plane with declination south of $< 30^{\circ}$ and right ascension range $>$7h and $<$21h. It also shows the number of pulsar imaging detections (column $\rm N_{\rm psr}$), the total number of pulsars in the 3dB (half power point) beam (column $\rm N_{\rm 3db}$) and the total number of sources (column $\rm N_{\rm src}$) extracted using AEGEAN for each Stokes I image before the application of any criteria.}
\label{table:obs}
\end{center}
\end{table}

The same data were also used to confirm image detections of known pulsars and follow up the candidates by beamforming the original VCS data and searching for pulsations. Using the Pawsey supercomputing systems (mainly the Garrawarla supercomputer dedicated to processing MWA data), processing an hour of observation to produce full Stokes images takes about 12 hours. Forming a beam and searching for pulsations using the same one-hour data, for a single pointing (single source) takes about 5 hours. This demonstrates the large volume of data captured by VCS and the large amount of time and resources required to process all the observations.

\subsection{Imaging pipeline}
\label{subsec:imaging}

\begin{figure*}[hbt!]
    \centering
    \includegraphics[width=\textwidth]{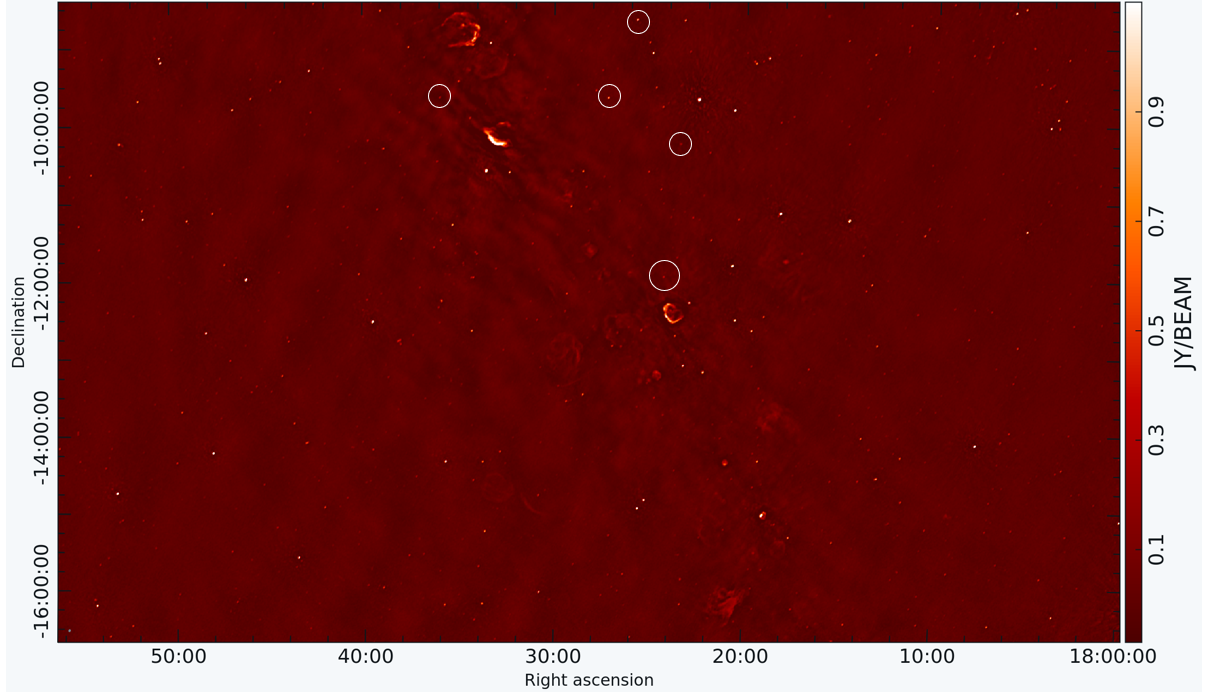}
    \caption[Stokes I image cutout of the on-Galactic Plane observation]{Stokes I image cutout of the on Galactic Plane observation. The pulsars detected in this part of the sky are circled in white. The mean RMS for the image is 5 mJy/beam. The Galactic supernova remnants can be seen in the middle of the image. Every single dot in the image corresponds to a source, the information of which is extracted using source extraction software, AEGEAN. The RMS of the image increases as we go closer to the Galactic Plane or at the image edges due to high confusion noise and lower sensitivity respectively. }
\label{fig:ongali}   
\end{figure*}

\begin{figure*}[hbt!]
    \centering
    \includegraphics[width=\textwidth]{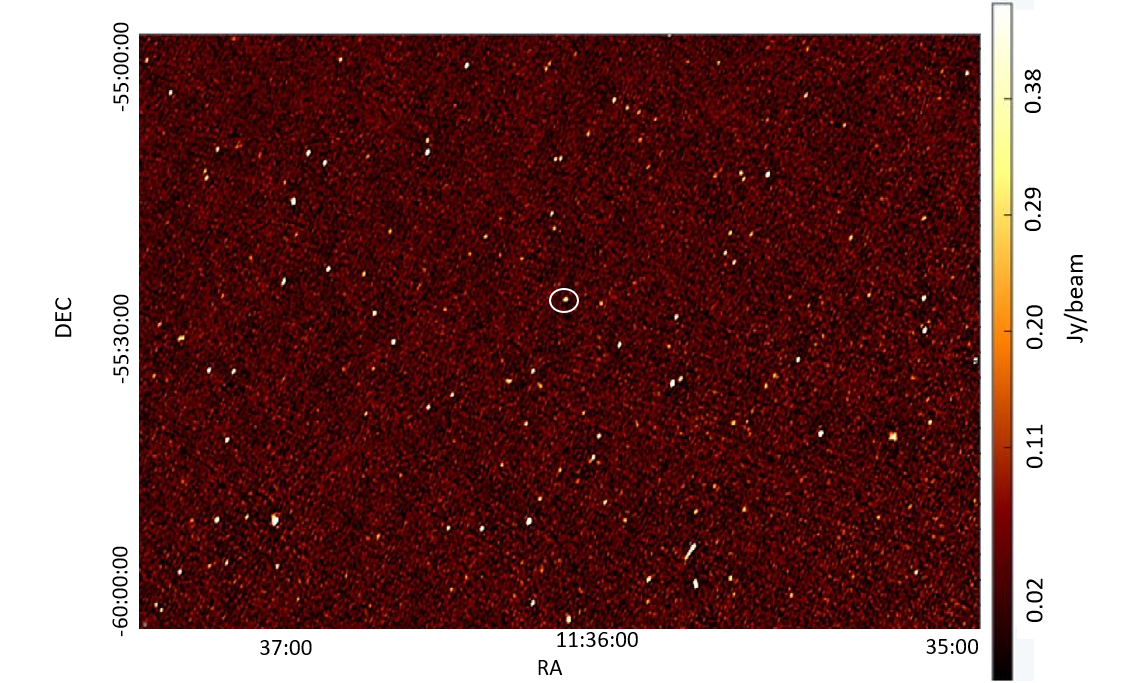}
    \caption[Stokes I image cutout of the off Galactic Plane observation]{Stokes I image cutout of the off Galactic Plane observation. The mean RMS for the image is 6 mJy/beam. As the Galactic Plane is not in the image, the RMS of noise is much lower due to significantly lower confusion noise. The number of sources in the off-GP image is lower, which makes the source finding and processing easier and faster. The pulsar in the field is circled in white. As this is an off-GP field the number density of pulsars is lower in comparison to an on-GP field.}
\label{fig:offgali}  
\end{figure*}

The raw voltages from the MWA antennas are processed by the xGPU software correlator \citep{ref:Ord} to produce visibilities with a temporal resolution of 1 second. The resulting data are then processed with COTTER \citep{ref:Off}, which converts them into the CASA measurement set format and applies calibration, and flags channels affected by radio-frequency interference (RFI) using the AOFlagger software \citep{ref:Offringa}. Calibration solutions are obtained from the MWA All-Sky Virtual Observatory \citep[MWA ASVO;][]{ref:Sokolowski}. Images in instrumental polarisation are created using WSCLEAN \citep{ref:wsclean} with a Briggs weighting of -1 and then transformed into Stokes I, Q, U, and V images using the MWA's "fully" embedded element beam model  \citep{ref:Soko}. The images are 8192$\times$8192 pixels with pixel size of 0.2\,arcmin, corresponding to $\sim$ 35\,$\degree$ $\times$ 35\,$\degree$ images. The individual 1-second images are averaged to produce mean full Stokes images, which are subsequently used for further analysis. An example of on and off GP Stokes I image are shown in Figure \ref{fig:ongali} and \ref{fig:offgali} respectively, where known detected pulsars are marked with white circles.

The source finding software, AEGEAN \citep[]{ref:Aegean1,ref:Aegean2}, was used to detect and extract radio sources from the mean Stokes I and V images. A catalogue of sources that exceeded a 5$\sigma$ threshold was created for each observation and analysed to generate a list of potential pulsar candidates. The pulsar candidates were chosen based on three criteria, namely, steep spectrum, circular polarisation and variability. Steep spectrum sources are the ones whose spectral index is steeper than -2. The spectral index is calculated using RACS flux and MWA flux of the source. The circular polarisation sources are the ones that are more than 7\% circularly polarised. The variable sources are selected by generating light curves for sources that have significance more than 5 and modulation of more than 20\%. More details of the methodologies can be found in \citet{ref:Sett}. The flux densities measured of detected pulsars and derived from Stokes I images were used for modelling their spectra using pulsar\_spectra software \citep{ref:psrspectra}, which is described in Section \ref{subsec:psrspectra}.

\subsection{Pulsar spectra fitting software - pulsar\_spectra}
\label{subsec:psrspectra}

Measurements of flux densities are important for detailed spectral analysis and furthering our understanding of pulsar radio luminosities. A recent version of the ATNF pulsar catalogue shows that the pulsar flux densities are well studied between 400\,MHz to 1.4\,GHz as most of the pulsars were discovered at those frequencies. However, they are not studied as extensively at frequencies $\leq$400\,MHz. The main challenges of accurate flux density measurements at these frequencies are pulsar scintillation. This is especially applicable for low-DM pulsars due to the interstellar medium. This leads to fluctuations in flux density from a factor of two to an order of magnitude. Even though pulsar scintillation is less prominent for high-DM pulsars, these would be very hard to detect at low-frequencies. Variations in flux densities are more dominant at low frequencies and is dependent on their Galactic latitude and instrumental parameters of the telescope such as observing bandwidth and integration time. The other factor that impedes reliable spectral models is that the currently catalogued data were taken with different instrumental backends and different systematic errors. Even though many accurate flux density measurements have been taken over the last several decades, there is no catalogue that records all this information. Furthermore, there is no theoretical model for the spectra of the pulsars and no single model that can accurately fit the wide variety of pulsars' spectra. 

\citet{ref:Jank} studied the spectral properties of 441 pulsars observed with the Parkes radio telescope at centre frequencies of 728, 1382 and 3100 MHz, providing a systematic and uniform sample of pulsar flux densities. The data were then combined with spectral data from the literature to derive the spectral properties of the pulsars and fit different spectral models to the combined data in a robust manner. The spectra of the pulsars needed to have at least four flux density measurements at four different frequencies to ensure sufficient spectral coverage and better constraints of the spectral fits. This simply means that it needs at least 4 or more data points to produce a fit as it would be more reliable than the one with two data points. The Akaike information criterion (AIC) was used to decide on the best-fit model. AIC measures the amount of information about the data retained by the model without over-fitting. The model that results in the lowest AIC was selected as the best-fitting model. 

Based on the method described in \citet{ref:Jank}, a spectral fitting software, \textbf{pulsar\_spectra} was developed by \citet{ref:psrspectra}. The \textbf{pulsar\_spectra} software contains an open-source catalogue of flux density measurements from several publications and uses this information to fit the best spectral model for a given pulsar. The \textbf{pulsar\_spectra} software also reduces the effect of underestimated uncertainties on outlier points and prevents the skewing of the model fit by using the Huber loss function \citep{ref:huber}. 

Currently, the software can incorporate different spectral models based on \citet{ref:Jank}. These are: (i) simple power law, (ii) broken power law, (iii) log parabolic spectrum, (iv) power law with low-frequency turnover, (v)power law with high-frequency cutoff and (vi) double turnover spectrum. The software is written in Python and can be easily installed. It has multiple features such as fitting a spectral model using the flux density measurements from the literature, estimating the flux density for a pulsar at a desired frequency and adding more spectral models to the repository.  Even though this is a relatively new software, it has already been applied to other analyses. One such spectral study was performed by \citet{ref:chris} for 22 radio pulsars detected with SKA-Low precursor stations. 21 out of the 22 pulsars showed a change in the spectral fit after the addition of the new low-frequency flux density measurements. A more extensive analysis using 893 radio pulsars is being undertaken by Swainston et al, 2023 (in prep).

The presented work uses the feature of \textbf{pulsar\_spectra}, which enables us to include the flux density measurements from this work and compare the spectral fits before and after the addition. In this study, pulsar mean flux density is measured from the continuum Stokes I images. The flux densities calculated from periodicity searches are dependant on beam model, sky model, sky and receiver temperatures. Given the dependencies on more models it is less reliable and may have large errors associated with it. The flux densities from the Stokes I images are reliant on proper calibration solutions and the beam model. Hence, the flux densities from Stokes I images therefore have fewer dependencies than that of the periodicity searches and therefore may have less errors associated with it. However, we do have to take into account that the flux densities from the Stokes I images may be overestimated due to the blending of sources and hence introduce errors to the measurements. Figure \ref{fig:sworklit} shows the comparison of the flux densities from this work to the literature and demonstrates that the flux densities from this work match well within the error limits of the flux densities in the literature. These low-frequency, reliable flux density measurements will be useful in filling up the gap in the measurements at the lower end of the frequency. Our study illustrates the scientific application of image-based pulsar surveys conducted using the Murchison Widefield Array (MWA) which leads to accurate measurements of low-frequency pulsar flux densities and broadband modelling of pulsar spectra. These insights serve as crucial inputs for informing both pulsar surveys and scientific endeavours planned with the Square Kilometer Array Low-Frequency (SKA-Low) telescope. The details of the results from this analysis are given in Section \ref{subsec:sed}. 

\section{Results and discussion}

\subsection{Detection of known pulsars}
\label{subsec:detect}

\begin{figure*}[hbt!]
    \centering
    \includegraphics[width=0.8\textwidth]{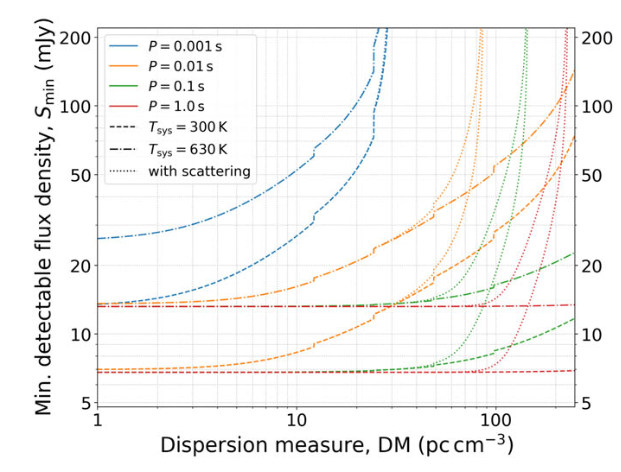}
    \caption[Minimum detectable flux density as a function of DM]{Minimum detectable flux density as a function of DM for the MWA SMART survey \citep{ref:smart1}. The same can be applied to this work due to similar observational and processing parameters. This shows the sensitivity limits for 10-min integration time for different pulse periods, for two different $T_{sys}$ (one for regions away from the Galactic Plane and the other for the mean in the plane excluding regions of Galactic Centre. The effect of pulse broadening is shown in the dotted lines \citep{ref:scintill}. The minimum pulsar flux density required for detection rapidly increases for pulsars beyond a DM of 250 $pc~cm^{-3}$ due to the significant effect of pulse broadening, making it harder to detect such pulsars in periodicity searches.}
    \label{fig:sdm}
\end{figure*}

Our image-based survey is generally expected to detect sufficiently bright pulsars irrespective of their DM. For MWA frequencies, the sensitivity for periodic searches significantly reduces above a DM of 250 pc $\rm cm ^{-3}$, but imaging sensitivity to pulsars is not compromised at these high DMs. Therefore, it is clear that there are three regions of DM-flux density parameter space: (i) at flux densities below the 5$\sigma$ imaging sensitivity of the telescope and low DM where only periodic searches are effective, (ii) at higher flux densities and DM $\leq$ 250 both image-based and periodic searches can be used, (iii) at higher flux and DM$\gtrsim$ 250 only image based searches are sensitive. This is summarised in Figure \ref{fig:sdm}.  

\begin{figure*}[hbt!]
    \centering
    \includegraphics[width=\textwidth]{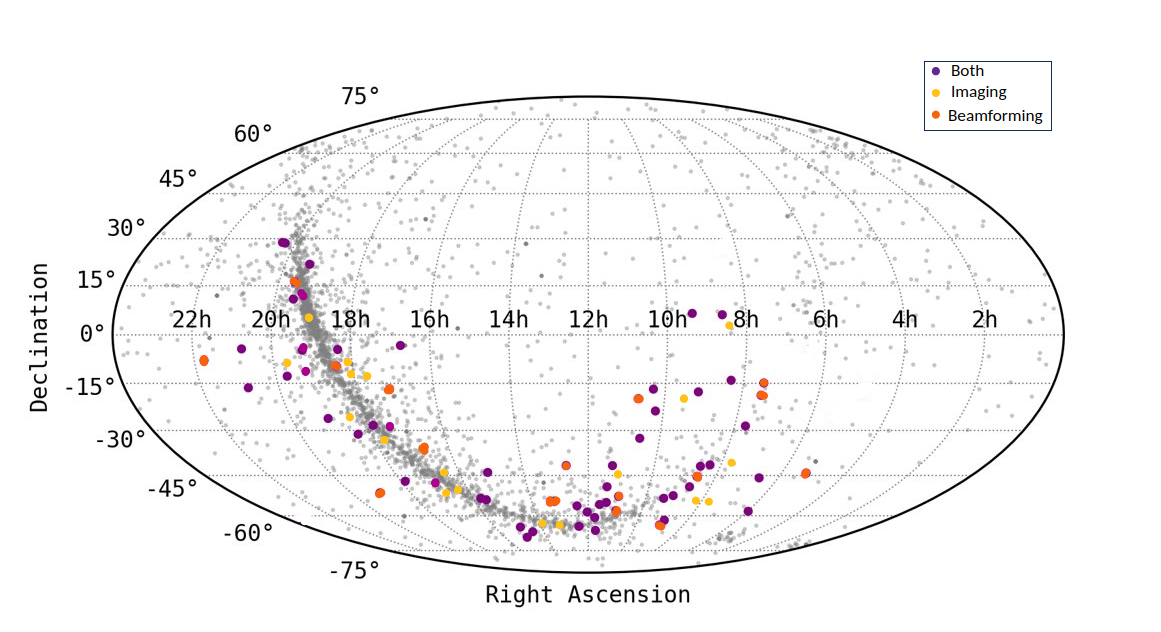}
    \caption[Distribution of known pulsars with colour dots showing pulsars detected in this work.]{Distribution of known pulsars with colour dots showing pulsars detected in this work. Grey-filled circles are the all-sky distribution of known pulsars in the ATNF pulsar catalogue \citep{ref:ATNF}. The purple-filled circles denote the pulsars that are detected by both periodicity searches and imaging. The yellow-filled circles are the pulsars that are detected only in imaging and the orange-filled circles are the pulsars detected only in periodicity searches.}
    \label{fig:psrdetect}
\end{figure*}

The Australian Telescope National Facility (ATNF) pulsar catalogue \citep[v1.70 \footnote{\url{https://www.atnf.csiro.au/research/pulsar/psrcat/}};][]{ref:ATNF} comprises of about 3000 pulsars detected to date. Of the total population, 1000 pulsars are present in the 3dB (half power point) for the 12 observations processed for this work. Taking the local $\sigma$ at the pulsar positions from Stokes I images and applying a flux density threshold for a 5$\sigma$ detection, we evaluate that 85\% (850) of the pulsars are below our detection threshold. As our work is focused on the dense region of the Galactic Plane, we also have to account for the pulsars that are present in supernova remnants (SNRs) which amounts to 4\% (40) of the pulsars, leaving 11\% (110) of the population which can be detected in image domain with the MWA Phase II sensitivity. Out of the 110 expected pulsar detections we were able to detect 66 pulsars in image-domain i.e. 60\% of the expected number of detections. The non-detections can be either due to spectral turnover of the pulsars at low frequencies \citep{ref:Jank} which leads to reduced flux density. The other possible reason can be that the pulsars are in regions of diffused emission such as in pulsar wind nebulae (PWN) or in undetected SNRs that are not accounted for and may be excluded as extended sources. Apart from detections of the pulsars in the image domain, time-domain beamformed searches resulted in 83 pulsar detections (17 more). The total number of pulsar detections in this survey is 83 shown in Figure \ref{fig:psrdetect}, including 3 MSPs and 4 binary pulsars. In Figure \ref{fig:psrdetect}, purple dots are the pulsars detected by both image-domain and time-domain searches, yellow dots are pulsars detected only in images and orange dots are the pulsars that are detected via periodic searches. For 14 pulsars, these are also the first low-frequency ($\leq$400 MHz) detections, and our flux density measurements can be helpful in more accurate spectral modelling of these pulsars. Some of the detections are discussed in detail in Section \ref{subsec:imagebeamform}.

\begin{figure*}[hbt!]
    \centering
    \includegraphics[width=\textwidth]{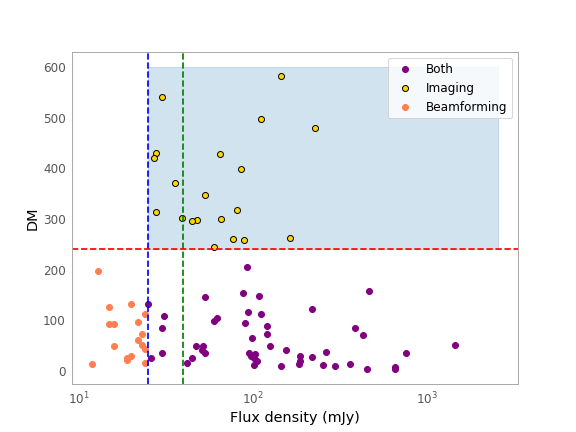}
    \caption[Known pulsars detected in this survey are shown in the DM-flux density plane.]{Known pulsars detected in this survey are shown in the DM-flux density plane. The purple-filled circles denote the pulsars that are detected by both periodicity searches and imaging. The yellow-filled circles are the pulsars that are detected only in imaging and the orange-filled circles are the pulsars detected only in periodicity searches. The blue and green dashed line indicate the mean flux density threshold for a 5$\sigma$ detection of pulsars in Stokes image of mean RMS of 5 mJy/beam and 8 mJy/beam respectively. The two thresholds are taken based on the range of mean RMS of the Stokes I image from the observations. The red dashed line indicates the DM threshold beyond which periodicity searches become less sensitive due to the scattering and DM smearing of the pulses at low frequencies. The blue-shaded region is the parameter space that is exclusively available to image-based searches. }
    \label{fig:timeimage}
\end{figure*}

Figure \ref{fig:timeimage} shows the pulsars detected on the DM-flux density plane. It is clear that the different search methods are optimal in different parts of the parameter space, and they overlap in the region of bright and low-DM pulsars. The purple dots denote the pulsars that are detected in both periodicity as well as imaging and as listed in a Table given in \citet{table:both}. These are mainly bright, low DM pulsars which have the highest overlap between the two search methods. The yellow dots denote the pulsars that are detected only in Stokes I images as listed in Table \citet{table:image}. These pulsars are mainly the high DM pulsars which are harder to detect via periodicity searches at low frequencies. The orange dots are the pulsars that are detected only in beamformed searches, the details of which are given in Table \citet{table:beam}. These pulsars are faint and hence are below the detection threshold of Stokes images. The blue-shaded region is the parameter space that is exclusively available to image-based searches. Image-based searches, therefore, can be useful in detecting high DM or highly scattered pulsars that may be missed by periodic searches at low frequencies but they are sensitive primarily to very bright sources. 

As the mean standard deviation of the noise ($\sigma$ or RMS) for these observations ranges from 5 mJy/beam to 8 mJy/beam, the blue and green dashed line in Figure \ref{fig:timeimage}, indicates a flux density threshold of 25 mJy/beam and 40 mJy/beam for the sensitivity of image-domain searches for a 5$\sigma$ detection of pulsars.

Obtaining a threshold for DM for periodicity searches is much more complicated as a DM cutoff is a complex relation between pulse broadening, frequency and period of the pulsar. Under the assumption of a pure Kolmogorov electron density spectrum \citep{ref:ism}, a non-linear function between scattering timescale, DM and frequency is given as $\tau_{d} \propto DM^{-2.2}\nu^{-4.4}$. This affects pulsar searches at low frequencies, especially when the pulse broadening time is longer than the pulsar's spin period, leading to a loss in sensitivity of periodicity searches. Figure \ref{fig:sdm} shows the significant effect of pulse broadening (i.e. smearing due to scattering) at larger DMs for the MWA SMART survey \citep{ref:smart1}. This is applicable to our work as we are using the same observing and processing parameters and observing frequency. The scatter broadening is also dependent on the line-of-sight and is larger in the Galactic Plane compared to off-Galactic Plane latitudes. Given that this work is focused on the Galactic Plane, pulse broadening needs to be taken into account when determining the DM cutoff. As shown in Figure \ref{fig:sdm}, the pulse broadening is higher at larger DMs. At higher DMs, the pulse broadening times are $\gtrsim$300 ms for a line of sight towards the Galactic Centre region at $|b|\lesssim 5^{\circ}$, $l\gtrsim 330^{\circ}$ or $l\lesssim 30^{\circ}$, where one would expect such high DMs. Furthermore, DM smearing is also significant at low frequencies (154\,MHz) and is $\sim$ 10\,ms at DM of $\sim$ 250\,$\rm pc~\rm cm^{-3}$ at 10kHz MWA VCS frequency channels \citep{ref:smart1}. Therefore, taking these effects into account we consider a virtual DM threshold of 250\,$\rm pc~\rm cm^{-3}$ beyond which DM smearing and pulse broadening significantly reduce the sensitivity of the search. This DM threshold is shown as a red dashed line in Figure \ref{fig:timeimage}.

\subsection{Pulsar detections - Imaging vs. Beamforming}
\label{subsec:imagebeamform}

\begin{figure}[hbt!]
    \centering
    \includegraphics[width=\textwidth]{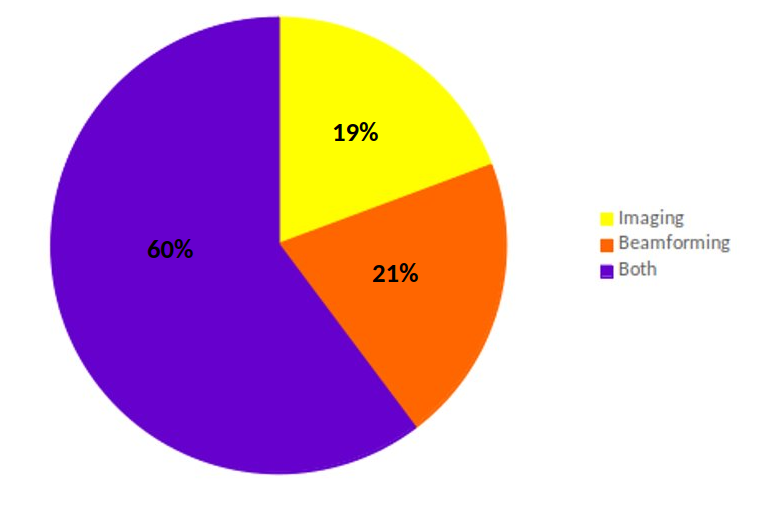}
    \caption[Percentages of the detections with the two search methods.]{Percentages of the detections with the two search methods. 60\% of pulsars were detected by both methods, while imaging is seen to perform as well as beamforming for the Galactic Plane observations.}
    \label{fig:ib}
\end{figure}

As mentioned earlier, both search methods are sensitive to certain areas of the DM-flux parameter space. Given that the total number of pulsar detections in this survey is 83, Figure \ref{fig:ib} shows the percentages of the detections with the two search methods. It shows that more than half of the pulsar detections were by both the search methods and image-based searches seems to perform as well as beamformed searches for the Galactic Plane observations processed as part of this work. 

\begin{figure*}[hbt!]
    \centering
    \includegraphics[width=\textwidth]{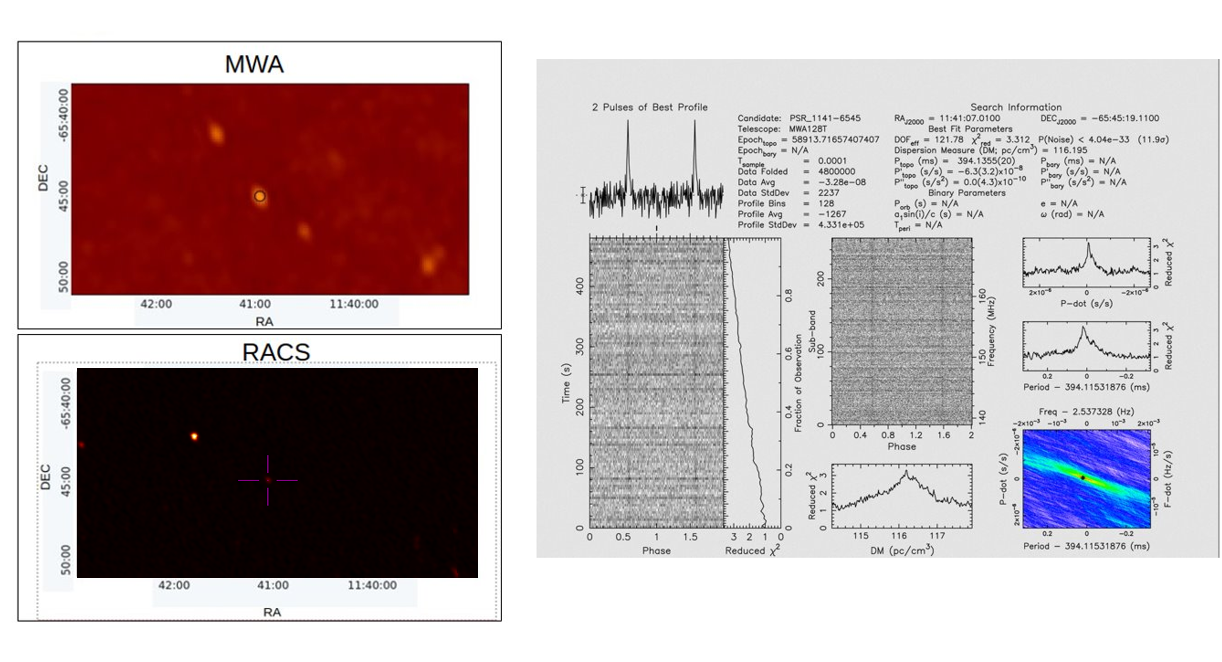}
    \caption[Detection of PSR J1141-6545 in both imaging and beamformed searches.]{Detection of PSR J1141-6545 in both imaging (13$\sigma$) and beamformed (12$\sigma$) searches. It shows the detection of the pulsar as a continuum source in MWA and RACS images (left panel) along with the PRESTO detection plot (right panel) based on the MWA data. The significance of the detection is comparable for both methods.}
    \label{fig:1141both}
\end{figure*}

60\% of the total pulsar detections were by both image-based as well as beamformed searches. An example of the detection of PSR J1141-6545 in both imaging and beamformed searches is shown in Figure \ref{fig:1141both}. It is seen as a 75 mJy (13 $\sigma$) source in imaging and a 12$\sigma$ PRESTO detection. The detection significance in both methods is comparable for this pulsar denoting that pulsars above our imaging sensitivity are mostly detected at similar significance with beamformed detections.  

\begin{figure*}[hbt!]
    \centering
    \includegraphics[width=0.6\textwidth]{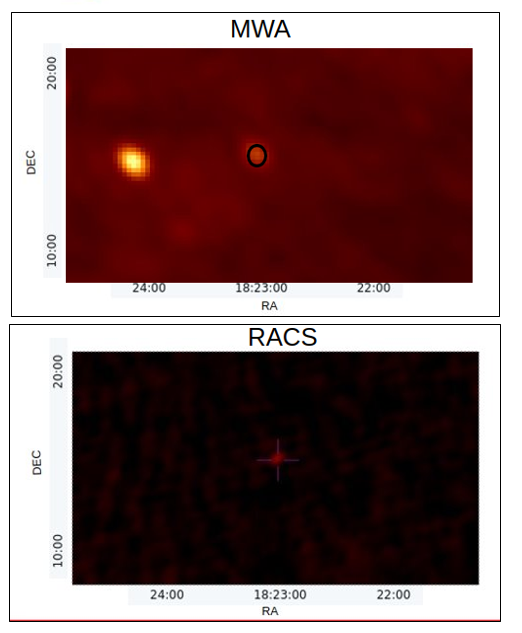}
    \caption[Image-based detection of PSR J1823-1115.]{Image-based detection of PSR J1823-1115. It can be seen as a 125 mJy continuum source in the MWA image (25$\sigma$). The corresponding source in the RACS Stokes I image is also shown in the bottom panel.}
    \label{fig:1823image}
\end{figure*}

\begin{figure*}[hbt!]
    \centering
    \includegraphics[width=\textwidth]{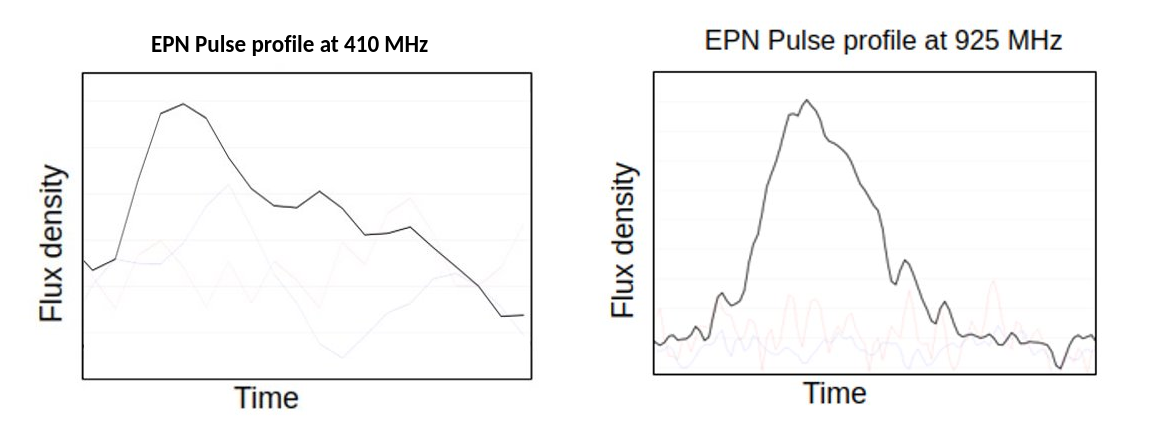}
    \caption[EPN pulse profile of PSR J1823-1115.]{EPN pulse profile of PSR J1823-1115 at 410 MHz and 925 MHz. It can be seen that the profile is more scattered at lower frequencies which makes it more difficult to detect at lower frequencies.}
    \label{fig:epn}
\end{figure*}

However, for the highly scattered or high DM pulsars, image-based searches are more favourable. One such example of PSR J1823$-$1115 with a high DM of 428.59 $\rm pc~\rm cm^{-3}$ is shown in Figure \ref{fig:1823image}. It shows the image-based detection of the pulsar in MWA as a 125 mJy (25$\sigma$) continuum source at the position of the pulsar as well as a detection in RACS Stokes I image. The beamformed search, however, did not result in any detection. The reason for the non-detection is possibly due to the highly scattered pulse profile of the pulsar at lower frequencies when compared to higher frequencies. This scattered profile is demonstrated in Figure \ref{fig:epn}, which shows the available profiles of the pulsar at 410 MHz and 925 MHz, in the European Pulsar Network (EPN \href{http://www.epta.eu.org/epndb/}) database. Due to the frequency dependence of scattering, we can expect the profile of this pulsar to be even more scattered at MWA frequency (185 MHz), making it more difficult to detect via beamformed searches. Periodicity searches at higher frequencies, such as that of RACS (888 MHz) would be able to detect it more comfortably. This demonstrates the capability of image-based searches at low frequencies to detect similar pulsars having scatter-broadened profiles which may be missed by periodicity searches.

\begin{figure*}[hbt!]
    \centering
    \includegraphics[width=\textwidth]{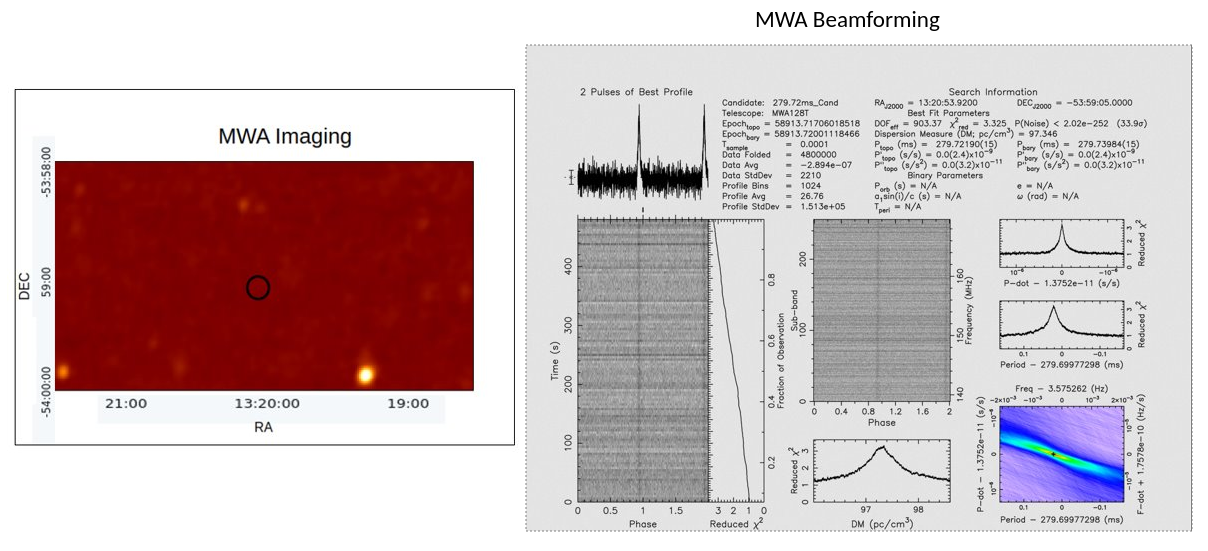}
    \caption[Beamformed detection of PSR J1320-5359.]{Beamformed detection of PSR J1320-5359 (right panel). The pulsar was below the detection threshold for the MWA Stokes I image and hence is not detected in the image domain (left panel).}
    \label{fig:1320beam}
\end{figure*}

Even though image-based searches are good for detecting high DM or highly scattered pulsars, the sensitivity of Stokes I images is still a limiting factor. This can be demonstrated by the example of detection of PSR J1320-5359 shown in Figure \ref{fig:1320beam}. This shows that the pulsar was detected at a high PRESTO detection significance of 33$\sigma$ but was not seen as a continuum source in the MWA Stokes I image. The expected mean flux density of the pulsar at a frequency of 184 MHz is $\sim$ 15 mJy/beam, which is below our best 5$\sigma$ detection threshold of 25 mJy/beam. Therefore, the sensitivity and confusion noise of MWA is a limiting factor in detecting fainter pulsars in image domain searches. However, the SKA-Low with approximately 64 more antennas will be a significantly more sensitive for image-based pulsar searches at low frequencies. The spatial resolution of SKA will also be much better than the present generation of telescope. The improved confusion noise and spatial resolution in the next-generation telescope compared to current facilities will lead to the detection of more pulsars in the image-plane.

\subsection{Comparison with literature}

\begin{figure*}[hbt!]
    \centering
    \includegraphics[width=\textwidth]{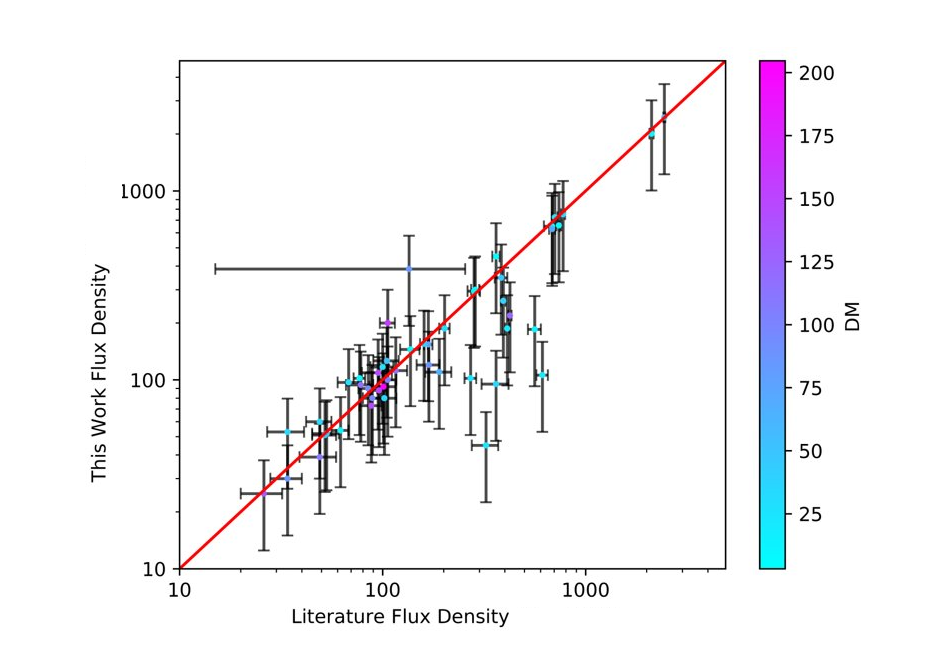}
    \caption[Pulsar flux density comparison of literature and this work.]{The comparison of pulsar flux density measurements in this work with the literature. All the pulsars detected via image-based method (66) are included in this plot. A 1:1 line is shown in red and most of the data points are within error limits from the line. The disagreements may be due to scintillation of the pulsars at low frequencies or differences in the mean flux density measurement method.}
    \label{fig:sworklit}
\end{figure*}

Figure \ref{fig:sworklit} shows the flux densities of the pulsars detected in the image plane as part of this work compared to their flux densities from the literature. Sixty-eight of the detected pulsars in the sample that is detected have previously measured flux densities at low frequencies. The surveys by \citet{ref:Murphy}, \citet{ref:Xue} and \citet{ref:smart2} provide a good comparison as they were also performed with the MWA, while \citet{ref:Frail1} conducted the survey using the GMRT at 150 MHz which aligns well with our MWA band of 184 MHz. 

Table \citet{table:both} shows the flux density of the pulsars from Stokes I images that are detected in both imaging and beamforming searches and the recorded low-frequency fluxes from the literature. Eighteen pulsars were previously detected by \citet{ref:Murphy} using MWA images generated as part of the GLEAM survey \citep[][]{ref:gl,ref:GLEAM}, with the GLEAM sub-band centred at 151 MHz, which matches well with our central frequency of 185 MHz. 24 pulsars were previously detected in \citet{ref:Xue} and six in \citet{ref:smart2} using periodicity searches. One pulsar was previously detected by \citet{ref:kondra} using LOFAR observations and periodicity searches. For the pulsars detected only in the Stokes I image, shown in Table \citet{table:image}, only two pulsars were previously detected using the GMRT 150 MHz radio continuum survey \citep[TGSS ADR,][]{ref:Frail1}. As can be seen in Figure \ref{fig:sworklit}, the flux densities measured in our work agree (within the errors) with these earlier surveys.  However, it is important to note that the fluxes from the literature are from different instruments and observational setups. The low-frequency measurements are also more affected by scintillation than higher-frequency measurements and pulsars are also intrinsically variable which may lead to differences in measured flux densities on particular days. The measured flux densities are also dependant on the DM of the pulsars and the corresponding scintillation effects it experiences. Moreover, flux densities may also vary depending on the detection method i.e. time domain detection using the radiometer equation to derive the flux density vs. image-based detections using the flux density of the sources directly from images. Hence, flux densities measured by different surveys and using different search methods may vary by an order of magnitude. Given this, the agreement of our flux density measurements with the earlier measurements (as shown in Figure \ref{fig:sworklit}, left panel) is remarkably good, and shows that MWA images can provide reliable flux density measurements despite being potentially prone to blending (caused by the spatial resolution of the MWA Phase II limited to $\sim$1\,arcmin). 

\subsection{Efficiency of the criteria}
\label{subsec:eff}

The pulsars detected in the image domain can be used as an efficiency marker for the developed criteria. It can be done by comparing the number of pulsars and candidates selected based on the threshold of each criterion. This method can help us understand the balance between actual pulsar detections and the number of false positives we get and improve the criteria such that we do not have numerous false positives when compared to actual detections. 

Table \ref{table:befcomb} shows the number of candidates and the pulsars that were selected by each criterion before the criteria were combined. The table also shows the efficiency of each criterion when the number of candidates and pulsars are compared. The spectral index criterion works the best when compared to the other two criteria, detecting 50 pulsars and a manageable number of candidates (less than 1000) for follow-up. The circular polarisation criterion was able to detect six pulsars as candidates out of 150 candidates and therefore shows the potential of circular polarisation surveys to detect pulsars that may be missed by other criteria. The variability criterion was able to detect two pulsars amongst 200 candidates. There are possible reasons for the low efficiency of the variability criterion. One of them can be that the pulsars detected in this survey do not vary on the timescales probed by our variability criterion. As diffractive scintillation is the main cause of variability of the pulsars in our timescale of minutes it is most applicable to low-DM pulsars. Given that this work is focused on the GP, it is unlikely that the variability criterion will be performing to its full potential due to the dominant population of high-DM pulsars (order of 100s $pc\, cm^{-3}$ in the Galactic Plane). 

In order to further reduce the number of candidates as well as to select sources that are the most probable pulsar candidates we attempt to combine the criteria and produce a list of candidates that satisfy more than one criteria. Table \ref{table:aftercomb} shows the candidates and the pulsars detected when the criteria are combined. The combination results in a reduction in the number of candidates. Four pulsars were detected by the combined criteria of steep spectrum and circular polarisation and two with the combined criteria of steep spectrum and variability. The other combinations do not yield any pulsar detections. Even though the combination of the criteria reduces the number of candidates, one has to be careful while combining the criteria. This is because it may happen that the combination of the criteria can excise good pulsar candidates that do not satisfy more than one criterion. This discrepancy in selecting candidates when combined criteria are applied arises mainly from the less understanding of the criteria. For example, the circular polarisation criteria may not be extremely reliable due to less reliable beam model leading to inadequate leakage characterisation. 

Overall, we conclude that the steep spectrum criterion is better than the other two when only individual criteria are used (not combined). Circular polarisation has its own advantage of being less affected by confusion noise and can be useful in detecting pulsars that may be missed in Stokes I image due to increased noise. Circular polarisation also has fewer number of candidates which makes it easier for follow up and confirmation. However, not all pulsars have circular polarisation and for some of the pulsars, the circular polarisation averaged over pulse period is zero due to sign flip of the Stokes V pulse profile. Moreover, it should also be noted that not all circularly polarised sources are pulsars. Variability criterion will be most fruitful if used for detecting low DM pulsars that are affected by diffractive scintillation. When criteria are combined, the combination of steep spectrum and circular polarisation works better than the other combination and has the highest efficiency. In an ideal case, one would follow up on all candidates, but due to the constraints of telescope time and computational costs, combining the criteria to rank the candidates is the most feasible way to reduce the number of follow-up candidates and select the best possible candidates for future work.

\begin{table}[H]
\begin{center}
\resizebox{\textwidth}{!}{%
\begin{tabular}{cccc}
\hline
Criterion      & Candidates & \begin{tabular}[c]{@{}l@{}}Pulsars amongst\\ candidates\end{tabular} & \begin{tabular}[c]{@{}l@{}}Efficiency\\ \%\end{tabular} \\
\hline \hline
Spectral index & 1000       & 50                                                                   & 5                                                       \\
Circular polarisation            & 150        & 6                                                                    & 4                                                       \\
Variability    & 200        & 2                                                                    & 1    \\                                                  
\hline
\end{tabular}% 
}
\caption[Table showing the efficiency of the criteria before they were combined]{Table showing the efficiency of the criteria before they were combined. It shows the candidates that each criterion selects and the corresponding pulsars that are selected. The last column shows the efficiency percentage for each of the criteria. Spectral index criteria perform better than the other two criteria.}
\label{table:befcomb}
\end{center}
\end{table}

\begin{table}[H]
\begin{center}
\resizebox{\textwidth}{!}{%
\begin{tabular}{cccc}
\hline
Combined Criteria     & Candidates & \begin{tabular}[c]{@{}l@{}}Pulsars amongst\\ candidates\end{tabular} & \begin{tabular}[c]{@{}l@{}}Efficiency\\ \%\end{tabular} \\
\hline \hline
\begin{tabular}[c]{@{}l@{}}Spectral index +\\ Circular Polarisation\end{tabular}          & 45         & 4                                                                    & 8                                                       \\
\begin{tabular}[c]{@{}l@{}}Spectral index +\\ Variability\end{tabular} & 100        & 2                                                                    & 2                                                       \\
\begin{tabular}[c]{@{}l@{}}Circular Polarisation +\\ Variability\end{tabular}             & 15         & -                                                                    & -   \\                                            
\hline
\end{tabular}% 
}
\caption[Table showing the efficiency of the combined criteria.]{Table showing the efficiency of the combined criteria. It shows the candidates selected by each combined criterion along with the corresponding pulsars detected. The final column indicates the efficiency percentage for each criterion. }
\label{table:aftercomb}
\end{center}
\end{table}

\subsection{Spectral energy distributions}
\label{subsec:sed}
The exact nature of the pulsar emission is an open question even after decades of research. Studying the radio spectra of the pulsars provides clues to their emission mechanism. However, these investigations are limited by the small number of flux density measurements, especially at frequencies below 400\,MHz, which limits the accuracy of spectral modelling. The flux density measurements, especially the first low-frequency detections of pulsars in this work will be helpful in bridging the gap between the low and high-frequency flux density measurements and help with broadband modelling of their spectra. 

For the 66 pulsars that are detected in the image-plane accurate flux density measurements can be obtained with 10\% - 20\% uncertainty depending on the position of the pulsar in the Stokes I image. Using the flux density measurements, we fitted spectra using the method implemented in the software, \textbf{pulsar\_spectra} \footnote{\url{https://github.com/NickSwainston/pulsar_spectra}} \citep{ref:psrspectra}. We attempted to investigate the change in the resulting spectral model, if any, upon addition of our low-frequency flux density measurements. The list of spectral fits used and the acronyms are given in Table \ref{table:acro}

\begin{table}[t]
\begin{center}
\resizebox{0.5\textwidth}{!}{%
\begin{tabular}{ccc}
\hline
Spectral model            & Acronym \\
\hline \hline
Simple power law          & SPL     \\
Broken power law          & BPL     \\
Low frequency turn over   & LFTO    \\
Double turn over spectrum & DTOS    \\
High frequency cut off    & HFCO   \\
\hline
\end{tabular}% 
}
\caption[List of acronyms for the spectral fits used in spectral modelling analysis in this work ]{List of acronyms for the spectral fits used in spectral modelling analysis in this work.}
\label{table:acro}
\end{center}
\end{table}

With only two flux density measurements, two-point spectral index can be calculated analytically (without fitting). Therefore, in order to increase the reliability of the spectral fits produced, the software, \textbf{pulsar\_spectra}, requires at least three data points for a simple power law fit and four or more data points for other spectral models. The software does not produce any fits if this requirement is not satisfied. Taking this threshold into consideration, out of the 66 pulsars, 63 had sufficient number of flux density measurements in the literature to produce the spectral fits without the addition of our work, one did not have enough measurements to fit the model without our data points and two did not have a sufficient number of data points to create a spectral fit even after the addition of our work. The lack of measurement data at frequencies below 400 MHz shows the need for more low-frequency campaigns to fill the gap in the literature. The reliable flux density measurement of pulsars with the MWA below 300 MHz will be useful in better spectral modelling of the pulsars. This is exemplified by the fact that 15 out of the 66 pulsars that were detected in Stokes I images changed the spectral models after the addition of our measurements as shown in Table \ref{table:specfits}. The table also shows the Akaike Information Criteria (AIC), which is a measure of how much information the model retains without over-fitting. The model that results in the lowest AIC is the most likely to be a sufficiently accurate model, which does not over-fit the data. Three different values of AIC is given: one which corresponds to model which fits the literature data ($\rm AIC_{\rm oo}$), one for the model that fits the data after the addition of data from this work ($\rm AIC_{\rm nn}$) and one for the previous model when fit to the new data set ($\rm AIC_{\rm on}$). However, it is important to note that AIC comparisons should only be done between different models applied to the same datasets. This means that it makes sense to compare $\rm AIC_{\rm nn}$ with $\rm AIC_{\rm on}$, but it does not make sense to compare $\rm AIC_{\rm nn}$ and $\rm AIC_{\rm oo}$. The lowest AIC when comparing $\rm AIC_{\rm nn}$ and $\rm AIC_{\rm on}$ gives us the model that best fits the data after the addition of the flux densities from this work. For ten of the pulsars the new model is the preferred one, whereas the old model is preferred for five of the pulsars. The change in the value of AIC ranges from as small as $\sim$ 1 to as large as $\sim$ 85. Any reduction in AIC is a meaningful improvement in the model and therefore even a small change is still significant. A rule of thumb is that a differences in AIC of the order of 20\% suggest preference to the model with a lower AIC value, whereas smaller differences can be treated as indication but are not really able to distinguish different models \citep{ref:aic}. Table \ref{table:specfits} shows that there are some cases like for PSRs J1312-5402 or J1807-2715 where the new model is much more preferred than the old model i.e $\rm AIC_{\rm nn}$ is much lower then $\rm AIC_{\rm on}$. There is also one case for PSR J1456-6843 where the old model is much more preferred. However, in many cases the differences between $\rm AIC_{\rm nn}$ and $\rm AIC_{\rm on}$ are smaller than 20\% and should be treated as an indication only. Further improvement can be achieved by incorporating more data covering sparsely populated frequency ranges. This demonstrates lack of low-frequency flux densities in the literature causes large uncertainties and volatility of the spectral modelling (i.e. addition of a single data point can change the fitted spectral model).

\begin{table}[t]
\begin{center}
\resizebox{0.8\textwidth}{!}{%
\begin{tabular}{cccccc}
\hline
PSRJ Name  & Old model & $\rm AIC_{\rm oo}$ \footnote{old model old data} & New model & $\rm AIC_{\rm nn}$ \footnote{new model new data}& $\rm AIC_{\rm on}$ \footnote{old model new data}\\
\hline \hline
J0820-1350 & LFTO & 231.4 & DTOS & 249.2 & 247.6 \\
J0820-4114 & BPL  & 89.9  & LFTO & 110.6 & 115   \\
J0907-5157 & LFTO & 43.2  & BPL  & 42.2  & 44    \\
J1121-5444 & HFCO & 25.1  & LFTO & 21.8  & 25.4  \\
J1312-5402 & SPL  & 24.0  & LFTO & 21.8  & 107.6 \\
J1456-6843 & LFTO & 88.6  & BPL  & 122.8 & 96.7  \\
J1709-1640 & HFCO & 157.9 & BPL  & 107.7 & 118.4 \\
J1752-2806 & DTOS & 740.4 & LFTO & 143.5 & 131.5 \\
J1913-0440 & LFTO & 226.3 & BPL  & 228.8 & 282.8 \\
J1932+1059 & LFTO & 166.2 & DTOS & 187   & 158.2 \\
J2022+2854 & LFTO & 141.4 & BPL  & 105.2 & 106.2 \\
J2048-1616 & LFTO & 323.6 & HFCO & 295.2 & 185.1 \\
J1600-5044 & LFTO & 45.7  & BPL  & 41.8  & 46.8  \\
J1807-2715 & SPL  & 24.6  & HFCO & 24.3  & 45.4  \\
J1827-0958 & SPL  & 20.3  & HFCO & 20    & 29.1  \\

\hline
\end{tabular}% 
}
\caption[Pulsars whose spectral fits changed after the addition of the flux density measurements from this work]{The list of pulsars whose spectral fits changed after the addition of the flux density measurements from this work. The table shows the model that fits the different data sets and the corresponding AICs. AIC is mainly aimed at finding the ''minimal model'' (corresponding to the lowest AIC value) which describes the data without over-fitting sufficiently well. The table shows the model and the corresponding AIC ($\rm AIC_{\rm oo}$) for the fits to the flux density data points from the literature. New model is the model that best fits the data after the addition of the flux density data points from this work. $\rm AIC_{\rm nn}$ is AIC value for this model fit. $\rm AIC_{\rm on}$ is the AIC obtained when the old model is fit to the new data. SPL stands for a simple power law, BPL is a broken power law, LFTO is a power law with low-frequency turn-over, HFCO is a power law with a high-frequency cut-off and DTOS is a double turn-over spectrum.}
\label{table:specfits}
\end{center}
\end{table}

\begin{figure*}[ht]

  \begin{minipage}[b]{0.5\linewidth}
    \centering
    \includegraphics[width=\textwidth]{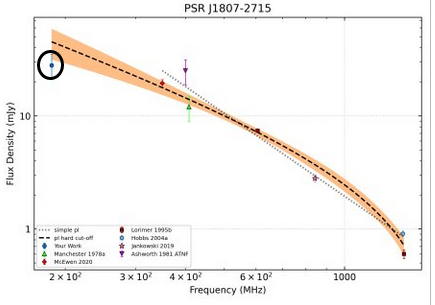} 
    %\caption{Initial condition} 
    \vspace{4ex}
  \end{minipage}%%
  \begin{minipage}[b]{0.5\linewidth}
    \centering
    \includegraphics[width=\linewidth]{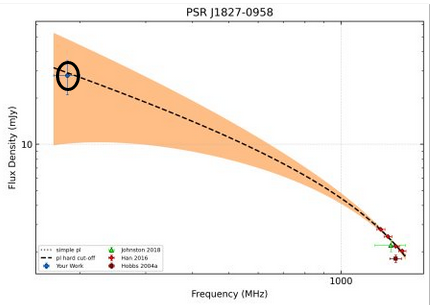} 
    %\caption{DFT, Initial condition} 
    \vspace{4ex}
  \end{minipage}%% 
\caption[Spectral fits for the pulsars that changed fits from SPL to HFCO.]{Spectral fits for the pulsars that changed fits from SPL to HFCO after the addition of our measurements. The flux density from this work is circled in black.}
\label{fig:spl_hfco}
\end{figure*}

\begin{figure*}[ht]

  \begin{minipage}[b]{0.5\linewidth}
    \centering
    \includegraphics[width=\textwidth]{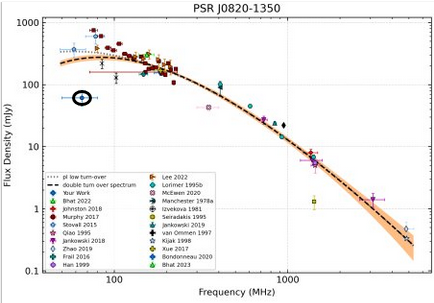} 
    %\caption{Initial condition} 
    \vspace{4ex}
  \end{minipage}%%
  \begin{minipage}[b]{0.5\linewidth}
    \centering
    \includegraphics[width=\linewidth]{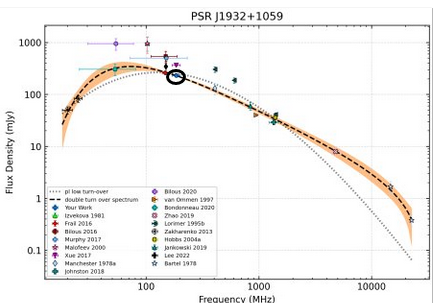} 
    %\caption{DFT, Initial condition} 
    \vspace{4ex}
  \end{minipage}%% 
\caption[Spectral fits for the pulsars that changed fits from LFTO to DTOS.]{Spectral fits for the pulsars that changed fits from LFTO to DTOS after the addition of our measurements. The flux density from this work is circled in black.}
\label{fig:lfto_dtos}
\end{figure*}

\begin{figure*}[ht]
\centering
    \includegraphics[width=0.5\textwidth]{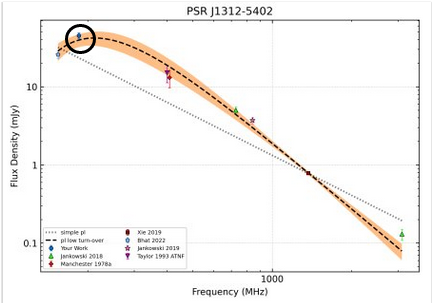}

\caption[Spectral fits for the pulsar that changed fits from SPL to LFTO]{Spectral fits for the pulsar that changed fits from SPL to LFTO after the addition of our measurements. The flux density from this work is circled in black.}
\label{fig:spl_lfto}
\end{figure*}

Five pulsars changed from LFTO to BPL (Figure \ref{fig:lpl_bpl}), two pulsars changed from SPL to HFCO (Figure \ref{fig:spl_hfco}), two pulsars changed from LFTO to DTOS (Figure \ref{fig:lfto_dtos}) and one pulsar changed from SPL to LFTO (Figure \ref{fig:spl_lfto}). The remaining five pulsars changed models to equally or more complex ones (Figure \ref{fig:others}). As we can see from the AIC values of the fit before and after the addition of the data from this work, the majority of the spectral fits after the addition of this work are favoured. For some of them, the values are comparable for both the fits. In a small number of cases, the AIC favours the model before the new data point addition.

\begin{figure*}[ht] 
 
  \begin{minipage}[b]{0.5\textwidth}
    \centering
    \includegraphics[width=\textwidth]{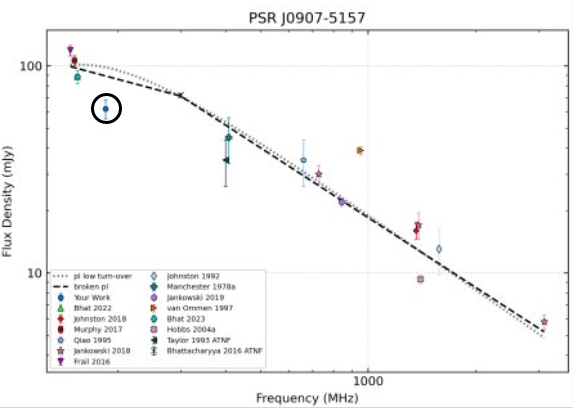} 
    %\caption{Initial condition} 
    \vspace{4ex}
  \end{minipage}%%
  \begin{minipage}[b]{0.5\linewidth}
    \centering
    \includegraphics[width=\textwidth]{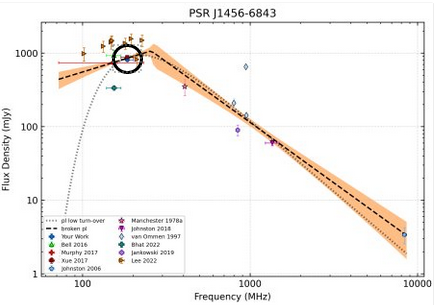} 
    %\caption{Rupture} 
    \vspace{4ex}
  \end{minipage}

  \begin{minipage}[b]{0.5\linewidth}
    \centering
    \includegraphics[width=\linewidth]{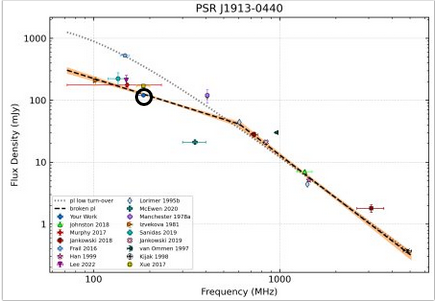} 
    %\caption{DFT, Initial condition} 
    \vspace{4ex}
  \end{minipage}%% 
  \begin{minipage}[b]{0.5\linewidth}
    \centering
    \includegraphics[width=\linewidth]{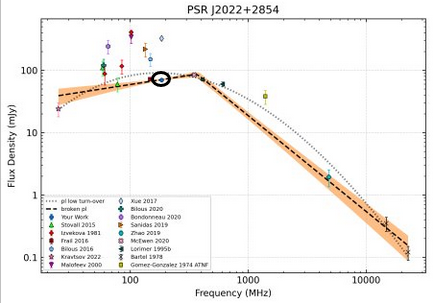} 
    %\caption{DFT, rupture} 
    \vspace{4ex}
  \end{minipage}
\caption[Spectral fits for the pulsars that changed fits from LFTO to BPL.]{Spectral fits for the pulsars that changed fits from LFTO to BPL after the addition of our measurements. The flux density from this work is circled in black.}
\label{fig:lpl_bpl}
\end{figure*}

\begin{figure}[ht]
\ContinuedFloat
\centering 
  
    \includegraphics[width=0.8\textwidth]{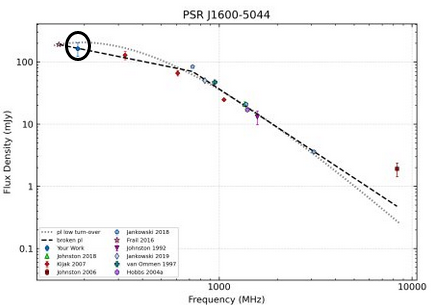} 
    %\caption{Initial condition} 

\caption[Spectral fits for the pulsars that changed fits from LFTO to BPL.]{Spectral fits for the pulsars that changed fits from LFTO to BPL after the addition of our measurements. The flux density from this work is circled in black.}

\end{figure}

\begin{figure*}[ht]

  \begin{minipage}[b]{0.5\linewidth}
    \centering
    \includegraphics[width=\textwidth]{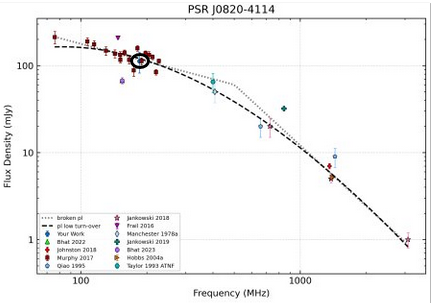} 
    %\caption{Initial condition} 
    \vspace{4ex}
  \end{minipage}%%
  \begin{minipage}[b]{0.5\linewidth}
    \centering
    \includegraphics[width=\textwidth]{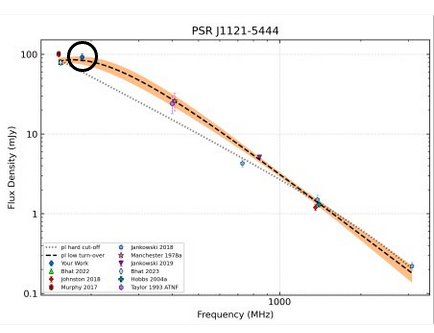} 
    %\caption{Rupture} 
    \vspace{4ex}
  \end{minipage} 
  \begin{minipage}[b]{0.5\linewidth}
    \centering
    \includegraphics[width=\linewidth]{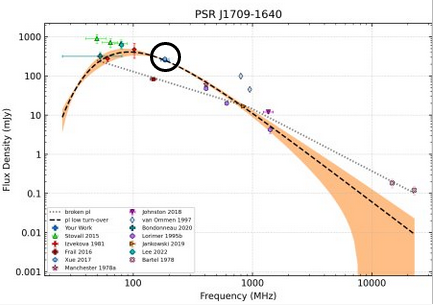} 
    %\caption{DFT, Initial condition} 
    \vspace{4ex}
  \end{minipage}%% 
\caption[Spectral fits for the pulsars that changed to other less or more complicated fits]{Spectral fits for the pulsars that changed to other less or more complicated fits after the addition of our measurements. The flux density from this work is circled in black.}
\label{fig:others}
\end{figure*}

\begin{figure*}[ht]
\ContinuedFloat
 
  \begin{minipage}[b]{0.5\linewidth}
    \centering
    \includegraphics[width=\textwidth]{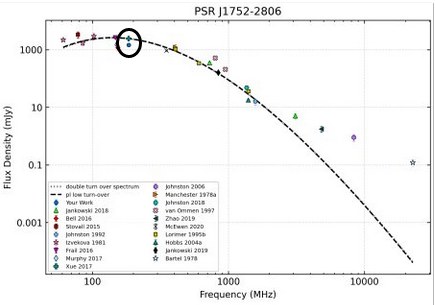} 
    %\caption{Initial condition} 
    \vspace{4ex}
  \end{minipage}%%
  \begin{minipage}[b]{0.5\linewidth}
    \centering
    \includegraphics[width=\textwidth]{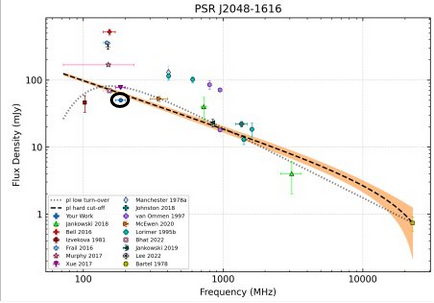} 
    %\caption{Rupture} 
    \vspace{4ex}
  \end{minipage} 
\caption[Spectral fits for the pulsars that changed to other less or more complicated fits]{Spectral fits for the pulsars that changed to other less or more complicated fits after the addition of our measurements. The flux density from this work is circled in black.}

\end{figure*}

\begin{figure*}[hbt!]
    \centering
    \includegraphics[width=0.5\textwidth]{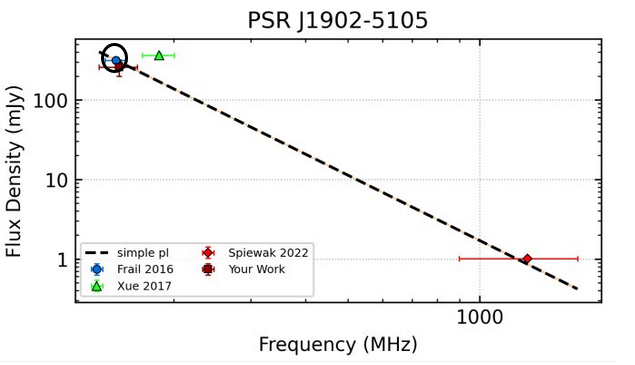}
    \caption[Simple power law fit for MSP J1902-5105]{Simple power law fit for MSP J1902-5105. The red box shows the low-frequency flux density measurement from this work. The flux density from this work is circled in black. }
    \label{fig:1902}
\end{figure*}

For millisecond pulsars, the broken power law for PSR J0737-3039A and double frequency turnover spectra for PSR J2145-0750 are consistent before and after the addition of our low-frequency flux density measurement. For one of the millisecond pulsars, PSR J1902-5105, a simple power law fit was produced (Figure \ref{fig:1902}) only after the addition of our flux density measurement. This shows the importance of performing more flux density measurements at lower and intermediate frequencies, which will improve the spectral modelling of pulsars and help explain the emission mechanism behind them. Measuring flux densities at multiple epochs is also important for improving the reliability of these measurements with reduced effects of flux variability. However, for two of the millisecond pulsars, PSR J1751-2737 and J1748-3009, we were unable to produce a reliable spectral fit even with the addition of our low-frequency measurements as \textbf{pulsar\_spectra} software requires at least three points for a simple power law fit and four or more for other spectral fits. 

\begin{figure*}[hbt!]
    \centering
    \includegraphics[width=0.5\textwidth]{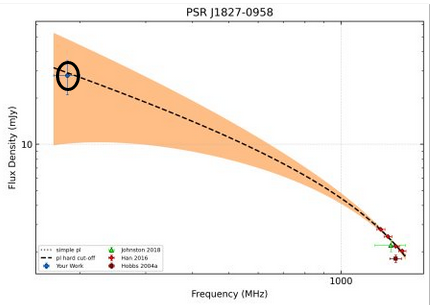}
    \caption[Broken power law fit for PSR J1827-0958]{Broken power law fit for PSR J1827-0958 after the addition of the low-frequency flux density measurements from this work. The spectral fit is less reliable due to a lack of measurements between the low and high-frequency measurements. The flux density from this work is circled in black.}
    \label{fig:1827}
\end{figure*}

Even though \textbf{pulsar\_spectra} provides reliable spectral fits, there can be discrepancies in the model fits due to several reasons. One of the reasons can be the lack or small number of measurements between 185 MHz and 1.4 GHz which may cause the fit to be less reliable. For example, the spectral fit for PSR J1827-0958 may be unreliable due to the lack of measurements between our flux density at 185 MHz and the measurements at higher frequencies as shown in Figure \ref{fig:1827}. The fits may also be skewed by underestimated error which can result from scintillation not being taken into account potentially increasing the measured flux density, especially at low frequencies. 

\subsection{The undetected pulsar population}
\label{subsec:nondetect}

Despite the detection of many known pulsars, there are many other which were not detected in imaging and periodic searches, that needs to be addressed. The non-detections in imaging can be attributed to either spectral turnover at low frequencies as highlighted by \citet{ref:Jank} or confusing the pulsars with other extended structures such as supernova remnants and pulsar wind nebulae near the Galactic Plane or the high noise confusion at the edge of the image. For periodic searches, the detection sensitivity is mainly affected by high scattering and scintillation which makes it harder to detect highly scattered pulsars and pulsars that tend to scintillate down making it fainter than the sensitivity threshold. 

\begin{figure*}[hbt!]
    \centering
    \includegraphics[width=0.5\textwidth]{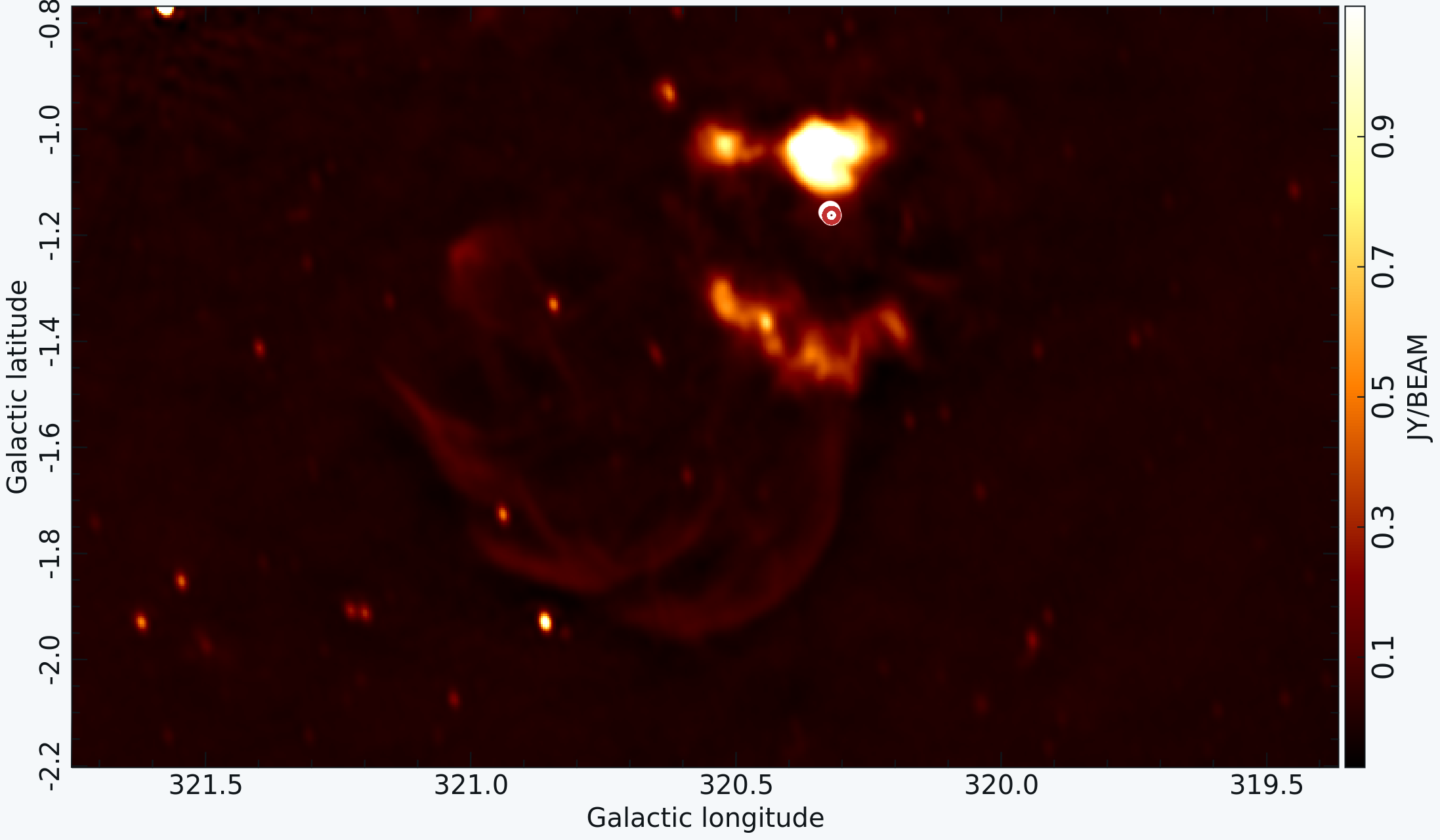}
    \caption[SNR G320.4-1.2 and PSR J1513-5908 is shown as a blue square amongst the SNR]{SNR G320.4-1.2 is shown in the middle of the image. The bright pulsar, PSR J1513-5908 is shown as a red dot amongst the diffuse emission of the supernova remnant. It is extremely difficult to detect this pulsar in the image domain due to the source being confused by the bright, extended emission of its surroundings. However, this can be detected in targeted beamformed searches.}
    \label{fig:snr}
\end{figure*}

The sky brightness temperature (hence sky noise) at low radio frequencies ($\sim$400 MHz) scales with frequency as $T_{b} \propto \nu^{-2.6}$ \citep{ref:haslam}, such that it becomes dominant at lower frequencies, making pulsar detections increasingly difficult. This affects both pulsed and imaging searches equally, with imaging searches also being affected by confusion and reduced image fidelity in the presence of bright Galactic HII regions and supernova remnants. The pulsar J1513-5908 is a good example of a bright pulsar confused by nearby bright, extended emission from supernova remnant, SNR G320.4-1.2 as shown in Figure \ref{fig:snr}.

The detectability at low frequencies is also affected by the spectral shape of the pulsar emission. Approximately 10\% of the known pulsar population is shown to have spectral turnovers \citep{ref:maron}. The pulsar population that we expect to detect also has a number of pulsars with known spectral turnovers such as PSR J1644-4559  which has a turnover at 600 MHz \citep{ref:Jank}. However, the lack of more accurate pulsar flux densities at low frequencies makes it difficult to confidently determine the spectral behaviour of the pulsars at frequencies below 300 MHz. 

For periodic searches, a factor that affects the detection of almost 40\% of the Southern pulsar population is the rapid decline in sensitivity beyond DM $\sim$ 250~ $\rm pc~ \rm cm^{-3}$ for time series data at MWA-frequencies. This is due to scattering and interstellar scintillation, which becomes very significant close to GP.

In summary, we conclude that the non-detections of the pulsars can be due to a combination of effects. There is clear evidence that the majority of the non-detections in the imaging plane are due to the increased confusion noise near the Galactic Plane and the low-frequency spectral turnover of the pulsars. The non-detections in periodic searches are a combination of factors such as reduced sensitivity at high DMs, scattering and scintillation.

\section{Summary and Conclusion}

The primary goal of the project was to discover new pulsar using the methodologies described in \citet{ref:Sett}. Even though we have not detected any new pulsars, we have detected multiple known pulsars leading to the most comprehensive imaging survey of Southern pulsars at low-frequencies with the MWA. We have identified 83 pulsars from the 12 MWA VCS observations covering the Galactic Plane. 66 of the pulsars are detected in wide-field MWA images. The approach of imaging is complementary to pulsation studies because it is not affected by pulse scatter broadening or dispersion , which potentially makes image-based searches sensitive to pulsars missed by  periodicity searches. This work also explores the unique parameter space that is exclusively accessible to image-domain pulsar searches eg., bright, high DM pulsars which are very difficult to detect at low frequencies using periodicity searches. We also presented the new low-frequency flux density measurements for 14 pulsars from the ATNF pulsar catalogue. These low-frequency flux density measurements can help to model pulsar spectra and address questions about the presence of low-frequency spectral turnovers. It is noticeable that with the scaricity of low-frequency flux density measurements of pulsars addition of a single point can drastically change the fitted models. Therefore, providing more flux density measurements at low-frequencies will improve the robustness of pulsar spectra modelling and lead to better understanding of pulsar emission mechanisms. 

We have carried out a preliminary spectral index study of our sample. We find that for 15 pulsars there is a change the spectral index model after the addition of our low-frequency measurement. However, these changes in the spectral index models may have been affected by the lack of measurements in intermediate frequencies such as 300 MHz or large fractional bandwidths of the existing measurements. Another factor that affects the spectral index model is the variability of the flux density measurements which may over or under-estimate the flux density measurement on a particular day. This work also emphasises the need for more flux density measurements at low frequencies. Especially, including more flux density measurements from different epochs, which will lead to better modelling and understanding of pulsar spectra and population of Southern sky pulsars. More flux density measurements can also answer questions about low-frequency spectral turnovers of pulsars.

Our identification of the pulsars was largely limited due to the sensitivity and the high confusion region near the Galactic Plane. The number of detections may also be affected by low-frequency spectral turnover for pulsars rendering them undetectable. We can conclude that a combination of all the effects is responsible for the pulsar non-detections in the image domain, while, time domain searches were mainly affected by scattering and scintillation. 

Based on the work done, we can conclude that imaging is a good way to find new pulsars in less explored parts of the DM space. Although, the main pulsar surveys will be conducted with the SKA-Mid, SKA-Low can also be very useful tool for searching for new pulsars mainly because of its larger FoV and many pulsars being brighter at lower frequencies. Image based strategies in the era of SKA may further expand the parameter space for these searches.

\begin{acknowledgement}
This scientific work uses data obtained from Inyarrimanha Ilgari Bundara / the Murchison Radio-astronomy Observatory. We acknowledge the Wajarri Yamaji People as the Traditional Owners and native title holders of the Observatory site. Support for the operation of the MWA is provided by the Australian Government (NCRIS), under a contract to Curtin University administered by Astronomy Australia Limited. We acknowledge the Pawsey Supercomputing Centre which is supported by the Western Australian and Australian Governments. Inyarrimanha Ilgari Bundara, the CSIRO Murchison Radio-astronomy Observatory and the Pawsey Supercomputing Research Centre are initiatives of the Australian Government, with support from the Government of Western Australia and the Science and Industry Endowment Fund. This research has made use of NASA's Astrophysics Data System. 
\end{acknowledgement}

% \paragraph{Funding Statement}

% This research was supported by grants from the <funder-name> <doi> (<award ID>); <funder-name> <doi> (<award ID>).

% \paragraph{Competing Interests}
% A statement about any financial, professional, contractual or personal relationships or situations that could be perceived to impact the presentation of the work --- or `None' if none exist.

\paragraph{Data Availability Statement}
The data can be made available on a reasonable request.

% A statement about how to access data, code and other materials allowing users to understand, verify and replicate findings --- e.g. Replication data and code can be found in Harvard Dataverse: \verb+\url{https://doi.org/link}+.

%\endnote in some journals will behave like \footnote; and \printendnotes will not output anything. 
\printendnotes

\printbibliography

% \appendix
% \section{Example Appendix Section}

\end{document}